\documentclass[aps,eqsecnum,preprint,prd,amsmath,amssymb,showpacs]{revtex4}

\usepackage{graphicx}
\usepackage{dcolumn}
\usepackage{bm}

\begin{document}
\addtolength{\jot}{10pt}

\def\bold#1{\setbox0=\hbox{$#1$}%
     \kern-.025em\copy0\kern-\wd0
     \kern.05em\%\baselineskip=18ptemptcopy0\kern-\wd0
     \kern-.025em\raise.0433em\box0 }
\def\slash#1{\setbox0=\hbox{$#1$}#1\hskip-\wd0\dimen0=5pt\advance
\dimen0 by-\ht0\advance\dimen0 by\dp0\lower0.5\dimen0\hbox
     to\wd0{\hss\sl/\/\hss}}
\def\ss{\langle \bar s s \rangle}
\def\qq{\langle \bar q q \rangle}
\def\mixedss{\langle \bar s \sigma g_s G s \rangle}
\newcommand{\be}{\begin{equation}}
\newcommand{\ee}{\end{equation}}
\newcommand{\bea}{\begin{eqnarray}}
\newcommand{\eea}{\end{eqnarray}}
\newcommand{\nn}{\nonumber}
\newcommand{\dd}{\displaystyle}
\newcommand{\bra}[1]{\left\langle #1 \right|}
\newcommand{\ket}[1]{\left| #1 \right\rangle}
\newcommand{\spur}[1]{\not\! #1 \,}
%
\vskip 1.0cm
\preprint{\vbox{\hbox{BARI-TH/05-510 \hfill}
                              }}
\vskip 2.0cm
\title{\bf
Radiative transitions  of  $D^*_{sJ}(2317)$ and $D_{sJ}(2460)$}
\author{P. Colangelo, F. De Fazio and A. Ozpineci \\}
\vskip 1.0cm
\affiliation{
Istituto Nazionale di Fisica Nucleare, Sezione di Bari, Italy\\}
\vspace*{1cm}
\begin{abstract}
\noindent
We study radiative decays   of  $D^*_{sJ}(2317)$ and $D_{sJ}(2460)$ using light-cone QCD sum rules. In particular, we consider the decay modes   $D^*_{sJ}(2317)\to D_s^* \gamma $
and $D_{sJ}(2460)\to D_s^{(*)} \gamma,  D^*_{sJ}(2317) \gamma$ and evaluate the hadronic parameters in the transition amplitudes analyzing  correlation functions of scalar, pseudoscalar,
vector and axial-vector quark currents. In the case of    $D^*_{sJ}(2317)\to D_s^* \gamma $ we also 
consider determinations based on two different correlation functions in HQET.  
The decay widths turn out to be different  than  previous estimates obtained by other methods;
 the results  favour  the  interpretation of   $D^*_{sJ}(2317)$ and $D_{sJ}(2460)$ as
ordinary $\bar c s$ mesons.
 
 \end{abstract}

\vskip 2.0cm
\maketitle

\clearpage

\section{Introduction} \label{sec:intro}

The observation of two narrow resonances with charm and strangeness,
$D_{sJ}^*(2317)$ in the  $D_s\pi^0$ invariant mass
distribution  \cite{Aubert:2003fg,Besson:2003cp,Bellecontinuo,BelledalB,Drutskoy,FOCUS2317,Aubert:2004pw}
and $D_{sJ}(2460)$  in the $D_s^* \pi^0$ and $D_s\gamma$ mass distributions
\cite{Besson:2003cp,Bellecontinuo,Aubert:2004pw,BaBar2460,BelledalB,BaBardalB},
has raised discussions about the nature of these states and their quark content
\cite{Colangelo:2004vu}.
The  natural identification  consists in considering these states  as the scalar and axial vector $\bar c  s$ mesons  respectively denoted as
$D_{s0}$ and $D_{s1}^\prime$.  In the heavy quark limit $m_c \to \infty$ such states are
expected to be degenerate in mass and to form a doublet having
$s_\ell^P=\displaystyle \frac{1}{2}^+$,  with $s_\ell$  the angular momentum of the light
degrees of freedom. In that interpretation the two mesons complete, together with  $D_{s1}(2536)$
and  $D_{s2}(2573)$, the set of  four states corresponding  to the
lowest lying P-wave $\bar c s$ states  of the constituent quark model.
A chiral symmetry between  the negative and positive parity doublets  $D_s-D_s^*$ {\it vs} $D_{s0} -D_{s1}^\prime$, suggested in ref.\cite{Bardeen,Nowak:2003ra}, would account for the  equality of the hyperfine splitting in the two doublets.

However, estimates of the masses of these mesons based on potential quark models generally  produce larger values  than the measured ones, implying  that the two scalar and axial-vector $\bar c s$  $D_{s0}$ and $D_{s1}^\prime$ states should be
heavy enough  to  decay  to $D K$ and $D^* K$ and   should have a  broad width.
 On this basis,  other interpretations  for $D_{sJ}^*(2317)$ and $D_{sJ}(2460)$ have been proposed,  for example  that of  molecular states
\cite{close}. Unitarity effects in the scalar $DK$ channel have also been considered
\cite{van beveren}.

Radiative transitions  probe the structure of hadrons, and therefore they are
suitable  to understand  the nature of $D_{sJ}^*(2317)$ and $D_{sJ}(2460)$ distinguishing among different 
interpretations  \cite{Godfrey:2003kg,Colangelo:2003vg}. Their rates
can be predicted by  various methods  and the predictions can be compared to the
experimental measurements.
 In particular, it has been suggested  that, in the molecular picture,
the $D_{sJ}(2460) \to D_{sJ}^*(2317) \gamma$ decay should be  driven by the $D^* \to D \gamma $ transition and  should occur at  a  different rate with  respect  to the rate for a  quark-antiquark meson decay
\cite{close};  such a suggestion has to be supported by  explicit calculations in view of the  experimental observations.

The radiative decay widths   
$D_{sJ}^*(2317) \to D_s^* \gamma$ and
$D_{sJ}(2460) \to D_s^{(*)} \gamma,  D_{sJ}^*(2317) \gamma$ have been evaluated using the 
constituent quark model \cite{Bardeen,Godfrey:2003kg} and the Vector Meson Dominance (VMD) ansatz in the heavy quark limit
\cite{Colangelo:2003vg,Colangelo:2004vu}. In this paper we use a different method, light-cone QCD sum rules, an approach exploited to analyze   many aspects of the heavy and light quark system phenomenology   \cite{altri,braun}, 
including radiative decays 
\cite{Balitsky:1989ry,Aliev:1995wi}
 (for a review  and references see \cite{Colangelo:2000dp}).
We  apply the method starting from the identification of $D_{sJ}^*(2317)$ with $D_{s0}$ and 
$D_{sJ}(2460)$ with $D^\prime_{s1}$ and
we  discuss  results and  related uncertainties.
In particular, in Section \ref{sec:Ds0} we consider the decay mode $D^*_{sJ}(2317) \to D_s^* \gamma$
and describe in detail  the calculation of the transition amplitude, the input quantities used in the analysis, the numerical results and the sources of uncertainties. 
In Section  \ref{sec:Ds0hq} we carry out a calculation of the same transition amplitude in the infinite heavy quark limit, discussing the deviation from the case of finite mass which is  sizeable  in the case of  charm.  The radiative modes of $D_{sJ}(2460)$ are analyzed in Sections
\ref{sec:Ds1a}-\ref{sec:Ds1c}, where we find different results with respect to those obtained by other methods. 
In Section \ref{sec:conc} we discuss the differences; in spite of them, considering the available experimental measurements, we conclude that the description of $D_{sJ}^*(2317)$ and
$D_{sJ}(2460)$ as $q \bar q$ states is favoured.
\section{$D^*_{sJ}(2317) \to D_s^* \gamma$} \label{sec:Ds0}

The amplitude of the E1 transition  $D_{s0} \to D^*_s \gamma$:
\be
\langle \gamma(q,\lambda) D_s^*(p,\lambda^\prime)| D_{s0}(p+q)\rangle =  e d  \left[ (\varepsilon^* \cdot \tilde \eta^*)(p\cdot q)-(\varepsilon^* \cdot p)(\tilde \eta^* \cdot q) \right]  \,\,\, , \label{eq:ampDs0}
\ee
with $\varepsilon(\lambda)$ and $\tilde \eta(\lambda^\prime)$ the photon and 
$D_s^*$ polarization vectors, respectively, and $e$ the electric charge, involves the hadronic
parameter $d$  which has dimension $\rm mass^{-1}$. According to the  strategy of QCD sum rules, 
the calculation of  this parameter starts from considering the QCD and the hadronic expressions
of  a suitable correlation function of quark currents.

We consider the correlation function
\be
F_\mu(p,q)=i \int d^4x \; e^{i p \cdot x} \langle \gamma(q,\lambda) | T[J^\dagger_\mu(x) J_0(0)] |0\rangle
\label{eq:corr-Ds0Ds*gamma}
\ee
of the scalar  $J_0=\bar c s$ and the vector $J_\mu=\bar c \gamma_\mu s$  quark currents, and an external photon state of momentum $q$ and helicity $\lambda$.  The correlation function  can be expressed
in terms of Lorentz invariant structures:
\be
F_\mu(p,q)= F_0 \;   \left[  (p \cdot \varepsilon^*)  q_\mu - (p\cdot q) \varepsilon^* _\mu \right]   + ... \,\,\ .
\ee

\begin{figure}[h]
 \begin{center}
\includegraphics[width=0.8\textwidth] {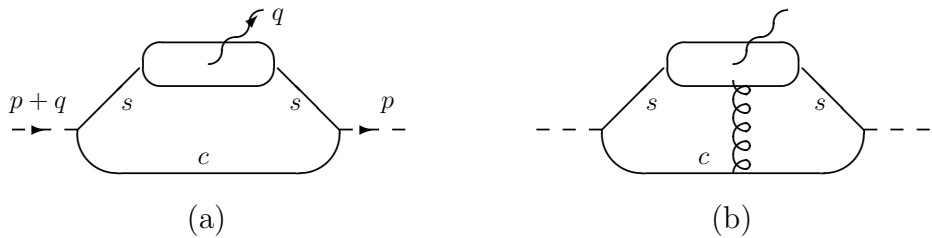}
\vspace*{0mm}
 \caption{\baselineskip=15pt Diagrams involving photon distribution amplitudes.
 The dashed lines represent the two currents in the correlation function
 (\ref{eq:corr-Ds0Ds*gamma}). In (a)   two-particle contributions and in (b) three-particle quark-gluon contributions are shown.}  
  \label{fig:light-cone}
 \end{center}
\end{figure}

In order to compute $F_0$ (or $F_\mu$) in QCD, we carry out  the light-cone expansion $x^2 \to 0$ of the product of the two
currents in (\ref{eq:corr-Ds0Ds*gamma}). This involves non-local matrix elements of quark operators
between the vacuum and the photon state which can be expressed in terms of operator matrix
elements of increasing twist.  For example, 
 contracting the charm-quark fields in eq.(\ref{eq:corr-Ds0Ds*gamma}) we obtain
\bea
F_\mu(p,q)&=& \int   \frac{d^4 k}{(2 \pi)^4}  \int d^4x \; \frac{e^{i (p -k) \cdot x}}{m_c^2 -k^2} \langle \gamma(q,\lambda) | \bar s(x) \gamma_\mu ( { \slash k} +m_c) s(0) |0\rangle \nn \\
&=&
\int \frac{d^4 k }{(2 \pi)^4} \int d^4x \; \frac{e^{i (p -k) \cdot x}}{m_c^2 -k^2} \Big[k_\mu
\langle \gamma(q,\lambda) | \bar s(x) s(0) |0\rangle \nn \\
&-&i k^\alpha
\langle \gamma(q,\lambda) | \bar s(x) \sigma_{\mu \alpha}s(0) |0\rangle+m_c
\langle \gamma(q,\lambda) | \bar s(x) \gamma_\mu s(0) |0\rangle \Big] ;
\eea
the expressions of   the photon matrix elements in terms of  distribution amplitudes are collected in the Appendix.  This kind of contributions  is depicted in fig.\ref{fig:light-cone}(a).  Moreover,
the light-cone expansion involves    higher-twist contributions  related to three-particle quark-gluon matrix elements, as  depicted in fig.\ref{fig:light-cone}(b); the expressions of the relevant  quark-gluon matrix elements  can also be found in the Appendix. 
%
%

In addition to the contributions of the photon emission from the soft $s$ quark, we  must consider
 the perturbative  photon coupling to the strange and charm quarks,
fig.2 (a) and (b).  It produces an expression for $F_0$ of the form:
\be
F_0=\int_{(m_s+m_c)^2}^{+\infty} ds \frac{\rho^P(s)}{(s-p^2)(s-(p+q)^2)} 
\ee 
\noindent
with
\bea
\rho^P(s)&=& -\frac{3 e_s}{4 \pi^2} \left\{-m_s \ln \left({s-m_c^2+m_s^2-\lambda^\frac{1}{2}(s, m_c^2,m_s^2) \over s-m_c^2+m_s^2+\lambda^{1\over 2}(s, m_c^2,m_s^2) } \right) +{m_c-m_s \over s}  \lambda^{1\over 2}(s, m_c^2,m_s^2) \right\} \nn\\
&+&{3 e_s \over 4 \pi^2} {m_c+m_s \over 2}  {\lambda^{1\over 2}(s, m_c^2,m_s^2) \over s} \left(1-{m_s^2-m_c^2 \over s } \right) \,\, + \,\,(s \leftrightarrow c)
\eea
($\lambda$ the triangular function).
Furthermore, nonperturbative effects when the photon is emitted from the heavy quark  give rise to contributions proportional to the strange quark condensate, corresponding to the diagram in fig.2 (c).

\begin{figure}[h]
 \begin{center}
\includegraphics[width=0.8\textwidth] {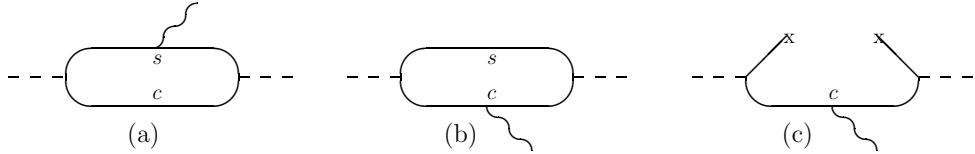}
\vspace*{0mm}
 \caption{\baselineskip=15pt  Perturbative photon emission by the strange (a) and  charm (b) quark.
 In (c) the strange quark condensate contribution is represented. }  
{} \end{center}\label{fig:pert}
\end{figure}
%
%
The result  is an expression of the correlation function
(\ref{eq:corr-Ds0Ds*gamma})  and of the function $F_0$ in terms of quantities such as quark masses, condensates and photon distribution amplitudes. The sum rule for $d$ is obtained by the equality
of this QCD expression with  a hadronic expression obtainted by a complete insertion of physical
states.
The two quark currents in (\ref{eq:corr-Ds0Ds*gamma})
 have non-vanishing matrix elements between the vacuum and  $D_s^*$ and $D_{s0}$:
\bea
\langle 0 | J_\mu^\dagger | D^*_s\rangle&=& f_{D_s^*} m_{D_s^*}  \tilde\eta_\mu \nn\\
\langle D_{s0} | J_0 | 0\rangle&=& f_{D_{s0}} m_{D_{s0}} 
\eea
\noindent
so that $F_\mu$ can be written as
\be
F_\mu = {\langle D_{s0} | J_0 | 0\rangle \langle \gamma D_s^*| D_{s0}\rangle \langle 0 | J_\mu^\dagger  | D_s^*\rangle\over (m^2_{D_{s0}}-(p+q)^2)(m^2_{D^*_s}-p^2)} +{\rm other \; resonances + continuum} \,\,\, ,
\ee 
neglecting the widths of $D_s^*$ and $D_{s0}$.
The sum rule  follows after a double Borel transformation in $-p^2$ and $-(p+q)^2$
of both the QCD and the hadronic representation of the correlation function, that  involves 
two Borel parameters,  $M_1^2$ and $M_2^2$. The transformation allows
to suppress the contribution of   the continuum of states and of  higher resonances, to suppress the higher twist terms in the QCD expression  of the correlation function and to 
remove all terms that  are either independent of one of the
two variables $-p^2$ or $-(p+q)^2$ or depend on it  only  polynomially. The  Borel parameters 
 $M_1^2$ and $M_2^2$ are independent;  we choose   $M_1^2=M_2^2$ since this allows,
invoking global quark-hadron duality  between the hadronic and the OPE
expression of the correlation function above some threshold $s_0$, to  subtract the continuum in the QCD side through  the substitution $e^{-{m_c^2 \over M^2}}\longrightarrow e^{-{m_c^2 \over M^2}}- e^{-{s_0 \over M^2}}$ in the leading twist term \cite{braun}. The masses of the charmed mesons involved in the transitions
are  close to each other, therefore the choice of equal Borel parameters is reasonable. The final expression of
 the sum rule for $d$ is:
\bea
d&=& \frac{e^{m^2_{D_{s0}}+m^2_{D^*_s}\over 2 M^2 }}{m_{D_{s0}} f_{D_{s0}}  m_{D^*_s} f_{D^*_s}} \Bigg\{ \int_{(m_c+m_s)^2}^{s_0} d s  \; e^{- {s \over M^2}} \rho^P(s) \nn\\
&+& e_c \; e^{-{m_c^2 \over M^2}} \ss \left(1+{m_s^2 \over 4 M^2} +{m_s^2 m_c^2 \over 2 M^4} \right)    
+ e_s \ss (e^{-{m_c^2 \over M^2}}- e^{-{s_0 \over M^2}}) M^2 \chi \phi_\gamma(u_0) \nn\\ 
&+& e_s \ss e^{-{m_c^2 \over M^2}}  \Bigg[ -\frac{1}{4} \left(\mathbb{A}(u_0)-8 \bar H_\gamma(u_0)\right)\left(1+{m_c^2\over M^2}\right)\nn\\
&+&\int_0^{1-u_0} dv \int_0^{u_0 \over 1-v} d \alpha_g  {\cal F}(u_0-(1-v)\alpha_g, 1-u_0-v \alpha_g,\alpha_g)\nn\\
&+&\int^1_{1-u_0} d v \int_0^{1-u_0 \over v} d \alpha_g  {\cal F}(u_0-(1-v)\alpha_g, 1-u_0-v \alpha_g,\alpha_g)\Bigg]\nn\\
&-& 2 e_s f_{3 \gamma} m_c e^{-{m_c^2 \over M^2}}  \Psi^v(u_0)) \Bigg\} 
\label{eq:srd}
\eea
where
${\cal F}= {\cal S} - \tilde{\cal S}-T_1+T_4-T_3+T_2+2 v (- {\cal S}+T_3-T_2)$,
$\dd \bar H_\gamma(u)=\int_0^u d u^\prime H_\gamma(u^\prime)$,
$\dd  H_\gamma(u)=\int_0^u d u^\prime h_\gamma(u^\prime)$ and
$\dd  \Psi^v(u)=\int_0^u d u^\prime \psi^v(u^\prime)$ . All the distribution amplitudes are defined  in the 
Appendix;  
$u_0={M_1^2 \over M_1^2+M_2^2}={1\over 2}$.

The sum rule (\ref{eq:srd}) involves the meson masses, for which we use the experimental data, and the
leptonic constants   $f_{D_s^*}$ and $f_{D_{s0}}$. For the former one, we put
 $f_{D_s^*}=f_{D_s}$ and use the central value of the experimental result
 $f_{D_s}=266\pm32$ MeV \cite{PDG}.  As for  $f_{D_{s0}}$,
a  sum rule  obtained from the analysis of the correlation function
\be
\Pi (p^2)=i \int d^4x \; e^{i p \cdot x} \langle 0 | T[J_0(0) J^\dagger_0(x)] |0\rangle,
\label{eq:corr-fDs0}
\ee  
\bea
 f_{D_{s0}}^2&=&\frac{e^{m^2_{D_{s0}}\over  M^2 }}{m_{D_{s0}}^2} \Bigg\{\frac{3}{8 \pi^2} \int_{(m_c+m_s)^2}^{s_0} ds \; \lambda^{1\over2}(s,m_c^2,m_s^2)\left[1-{(m_c+m_s)^2\over s}\right] e^{-{s\over M^2}} \nn\\
 &+&{ e^{-{m^2_c \over  M^2}}\over 2} \left[ \ss \left(2 m_c-m_s -{m_c^2 m_s \over M^2}+{m_c^3 m_s^2 \over M^4}\right)-{\mixedss\over 2}{m_c^3\over M^4}\right]\Bigg\}
\eea
allows to obtain $f_{D_{s0}}=225\pm25$ MeV, using the parameters  in the Appendix.

 From eq.(\ref{eq:srd}) we can compute $d$ varying the threshold $s_0$ and considering 
 the range of the external variable $M^2$ where the result  is independent on it (stability region).   
 In this region a hierarchy in the terms with increasing twist is observed,
 so that we can presume that the neglect of higher-twist contributions induces a small error. 
 On the other hand,  
 the perturbative term, which depends on both the light and the heavy quark charges,  represents
 a sizeable contribution to the sum rule.
 
\begin{figure}[h]
 \begin{center}
\includegraphics[width=0.5\textwidth] {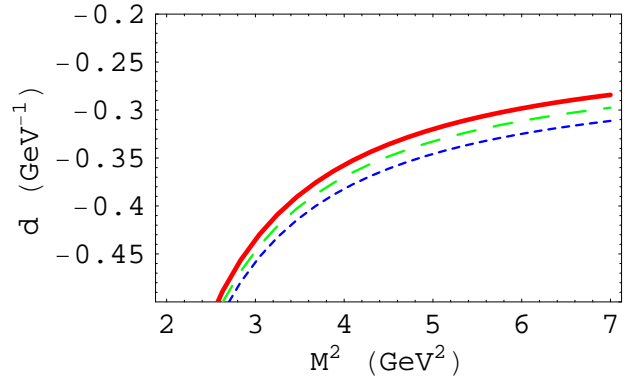}
\vspace*{0mm}
 \caption{\baselineskip=15pt 
The  parameter  $d$   in the  $D_{s0} \to D_s^* \gamma$ 
decay amplitude  eq.(\ref{eq:ampDs0}) versus the Borel parameter $M^2$ . The curves correspond to the thresholds $s_0=2.45^2$ ${\rm GeV}^2$ (continuous  line),  $s_0=2.5^2$ ${\rm GeV}^2$ (long-dashed line) and $s_0=2.55^2$ ${\rm GeV}^2$ (dashed line).}  
  \label{fig:d}
 \end{center}
\end{figure}

In fig.\ref{fig:d}  we plot  the curves corresponding  to different values of $s_0$.
%
Considering the range $5 \,\, {\rm  GeV}^2 \le M^2 \le 7 \,\,{\rm GeV}^2$,
where the best stability in  $M^2$ is found,  together with the variation of the threshold $s_0$, we get: $ -0.35 \,\, {\rm GeV}^{-1} \le d \le -0.28 \,\, {\rm GeV}^{-1}$, corresponding to the radiative decay width
\be
\Gamma(D_{s0}\to D_s^* \gamma)= \, (4\, - \, 6) \, \, {\rm keV}. \label{eq:rateDs0}
\ee

In  (\ref{eq:rateDs0}) we have  only considered the uncertainty in $s_0$ and $M^2$, and we have used the central values of the QCD parameters collected in the Appendix.  Actually, such parameters represent another source of  uncertainty. In particular,  an important input parameter is the magnetic susceptibility of the quark condensate,  $\chi$, for which we use the value  determined 
in ref.\cite{Ball:2002ps}:  $\chi=-(3.15 \pm 0.3) \,{\rm GeV}^{-2}$.
A  different value $\chi=-4.4 \, {\rm GeV}^{-2}$, previously used  in other  sum rule analyses, would produce a $40\%$  larger value of $ | d |$.

The result (\ref{eq:rateDs0}) shows that the radiative decay occurs at a typical rate for this  kind
of transitions (a few keV's).  However, the rate is larger by a factor of 4-5 than that obtained using VMD and the infinite heavy quark limit, and by a factor of 2-3 larger than
the estimates based on the constituent quark model.
 It is interesting to investigate the reason of the numerical differences;  aiming at that,
we  estimate  $d$ by  light-cone QCD sum rules  in the heavy quark limit,
 using an approach based on the heavy quark effective theory. We discuss such a calculation   
in the next Section.

\section{$D_{sJ}(2317) \to D_s^* \gamma$ in the heavy quark limit} \label{sec:Ds0hq}

In order to determine the hadronic parameter  $d$  in eq.(\ref{eq:ampDs0}) when $m_c \to \infty$, we consider two different correlation functions:
\be
F_\mu^{(S)}(\omega, q \cdot v)=i \int d^4x \; e^{i (\omega v -q) \cdot x} \langle \gamma(q,\lambda) | T[ \hat J^\dagger_\mu(x) \hat J_0(0)] |0\rangle
\label{eq:corr-Ds0Ds*gamma-hqs}
\ee
and
\be
F_\mu^{(D)}(\omega, q\cdot v)=i \int d^4x \; e^{i (\omega v -q) \cdot x} \langle \gamma(q,\lambda) | T[\hat J^\dagger_\mu(x) \hat J_d(0)] |0\rangle \,\,\, .
\label{eq:corr-Ds0Ds*gamma-hqd}
\ee
The currents in (\ref{eq:corr-Ds0Ds*gamma-hqs}) and (\ref{eq:corr-Ds0Ds*gamma-hqd}) are effective
currents constructed
in terms of the strange quark field and of the effective field $h_v$ of the heavy quark 
(in our case the charm quark) with four-velocity $v$. The effective field $h_v$ is related to
 the heavy quark field $Q$ in QCD through the relation
$\displaystyle h_v= e^{i m_Q v \cdot x} \frac{ 1+{\slash v}}{2} Q$ (for a review see  \cite{Neubert:1993mb}). 
The current
$\hat J_\mu= \bar h_v \gamma_\mu s$ has  the quantum numbers of a vector meson. On the other hand, the currents
$\hat J_0= \bar h_v  s$ and $\hat J_d= \bar h_v (-i) \gamma^\mu \vec D_{t \mu} s$
have both the quantum numbers of a scalar meson, since $D_{t \mu}\equiv g_{t \mu \alpha} D^\alpha\equiv (g_{\mu \alpha} - v_\mu v_\alpha)D^\alpha$,
$D$ being the covariant derivative. The latter current has been proposed as better suited
for describing scalar heavy-light quark mesons   in the heavy quark limit \cite{Zhu:1998ih},
therefore it is interesting to investigate how it  behaves in sum rules for radiative decays.

The sum rules for $d$ are obtained  from (\ref{eq:corr-Ds0Ds*gamma-hqs}) and (\ref{eq:corr-Ds0Ds*gamma-hqd}) using the same procedure followed in  Sec.\ref{sec:Ds0}, namely considering the light-cone expansion and the hadronic representation of the correlation functions, making a double Borel
transform in the variables $\omega$ and $\omega^\prime=\omega-q \cdot v$ that involve two Borel parameters $E_{1,2}$,
 choosing $E_1=E_2=2 E$ and invoking quark-hadron duality above some threshold $\nu_0$.   
From  (\ref{eq:corr-Ds0Ds*gamma-hqs}) we obtain \cite{comment}

\bea
d^{(S)} &=& \frac{4}{\hat F \hat F^+} e^{\bar \Lambda+\bar \Lambda^+ \over 2 E}
 \Bigg\{ \frac{3 m_s e_s}{8 \pi^2} \int_{m_s}^{\nu_0} d \nu  \; e^{- {\nu \over E}} 
 \ln \left[{\nu -(\nu^2-m_s^2)^{1\over2} \over \nu +(\nu^2-m_s^2)^{1\over2} }\right]  \nn\\
&+& e_s {\ss \over 2} E \chi \phi_\gamma(u_0)  \left(1-e^{-{\nu_0 \over E}}\right) -
e_s  \frac{\ss}{4 E} \left( \frac{\mathbb{A}(u_0)}{8}- \bar H_\gamma(u_0)\right) -
\frac{e_s f_{3\gamma}}{2}\Psi^v (u_0)\Bigg\} \,\,\, .  \nn \\
\label{eq:d1-hqet}
\eea
On the other hand, from the correlation function (\ref{eq:corr-Ds0Ds*gamma-hqd}) we get:
\bea
 d^{(D)} &=& \frac{4}{\hat F \hat F^+_d} e^{\bar \Lambda+\bar \Lambda^+ \over 2 E}
 \Bigg\{ -\frac{3 m_s e_s}{8 \pi^2} \int_{m_s}^{\nu_0} d \nu  \; e^{- {\nu \over E}} 
 \ln \left[{\nu -(\nu^2-m_s^2)^{1\over2} \over \nu +(\nu^2-m_s^2)^{1\over2} }\right](m_s+\nu)  \nn\\
&+& E   \frac{e_s f_{3\gamma}}{2} \left[ \Psi^v(u_0)+ \frac{1}{4} \psi^a(u_0)-u_0 \frac{\psi^{\prime a}(u_0)}{4}\right] \left(1-e^{-{\nu_0 \over E}}\right)\nn\\
&+&  e_s \ss \frac{1}{2} \Bigg[-E^2 \left(\chi \phi_\gamma(u_0) +u_0 \chi \phi^\prime_\gamma(u_0)\right)\Bigg ]\left(1-e^{-{\nu_0 \over E}}(1+\frac{\nu_0}{E})\right) \nn\\
&+&  e_s \ss \frac{1}{2} \Bigg[\frac{1}{16}\left(\mathbb{A}(u_0)+u_0 \mathbb{A}^\prime(u_0)\right)-
 \bar H_\gamma(u_0) \Bigg]  \Bigg\} \,\,\, .\nn \\ \label{eq:d1-hqet-1}
\eea
Notice that in (\ref{eq:d1-hqet})-(\ref{eq:d1-hqet-1})
only photon emission from the light quark contributes. 
In the heavy quark limit  the current-vacuum matrix elements are defined as follows:
$\langle 0 | \hat J^\mu | D_s^*(v,\lambda) \rangle_H= \hat F \tilde \eta^\mu(\lambda)$,
$\langle 0 | \hat J_0 | D_{s0}(v) \rangle_H= \hat F^+ $,
$\langle 0 | \hat J_d | D_{s0}(v) \rangle_H= \hat F^+_d $ (the subscript $H$ indicates that the states are normalized as used in HQET;  $\hat F^{(+)} $ and $\hat F^+_d $  have dimension $\rm mass^{3/2}$ and $\rm mass^{5/2}$, respectively).  Moreover,  
$\bar \Lambda$ and  $\bar \Lambda^+$ are mass  parameters defined as
$\bar \Lambda=m_{D_s^*}-m_c$, $\bar \Lambda^+=m_{D_{s0}}-m_c$
(in the heavy quark limit).
We use the  numerical values: $\hat F=0.35 \; {\rm GeV}^{\frac{3}{2}}$, $\hat F^+=0.45 \;{\rm GeV}^{\frac{3}{2}}$, $\hat F^+_d=0.44  \;{\rm GeV}^{\frac{5}{2}}$ and
$\bar \Lambda =0.5$ GeV, $\bar \Lambda^+ =0.86$ GeV   \cite{Neubert:1993mb,Colangelo:1995ph,Zhu:1998ih,dai}.
%
\begin{figure}[h]
 \begin{center}
  \includegraphics[width=0.4\textwidth] {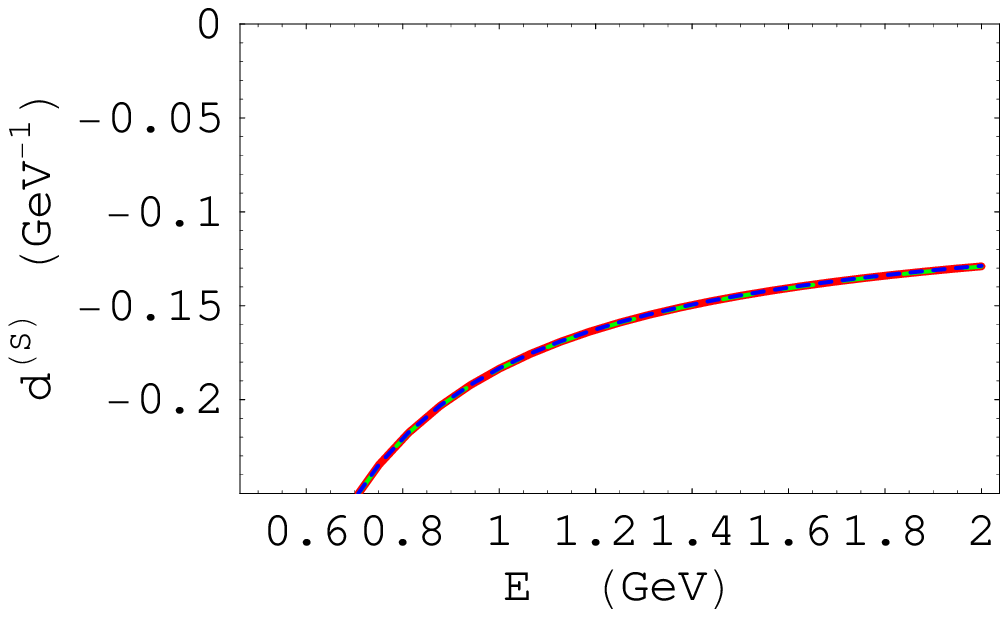}
  \hspace*{14mm}
  \includegraphics[width=0.4\textwidth] {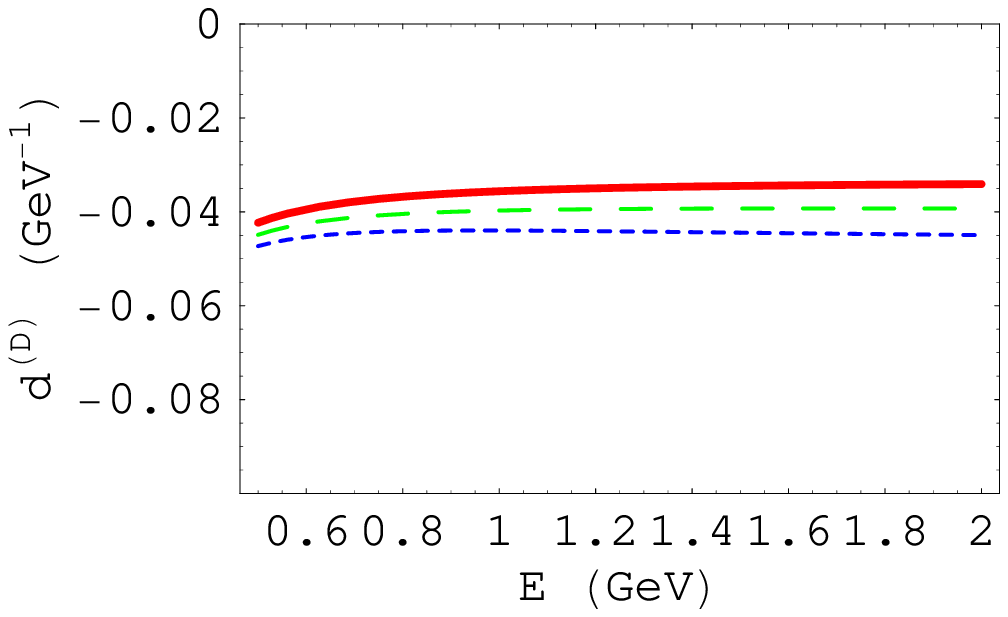}
\vspace*{0mm}
 \caption{\baselineskip=15pt 
 The parameters $ d^{(S)}$   obtained from  eq.(\ref{eq:d1-hqet}) (left) and $ d^{(D)}$ 
from   eq.(\ref{eq:d1-hqet-1})  (right) 
 versus the Borel parameter $E$. The  continuous, long-dashed  and dashed lines correspond to the thresholds $\nu_0=1.1, 1.2$ and $1.3$ GeV, respectively.}
  \label{fig:d1-hqet}
 \end{center}
\end{figure}
%
In fig.\ref{fig:d1-hqet} (left)  we depict the result corresponding to  eq.(\ref{eq:d1-hqet}). Considering  the region where  $d^{(S)}$ is independent of the Borel parameter $E$:
$1.2 \, {\rm GeV} \le E \le 1.6 \, {\rm GeV}$, and  the variation of the threshold $\nu_0$,  we obtain
$-0.16  {\rm GeV}^{-1} \le d^{(S)}\le -0.13$ GeV$^{-1}$. 
Therefore, we obtain  in the heavy quark  limit a value  compatible with the value obtained by VMD
in the same limit: $d \simeq  -0.15$ GeV$^{-1}$;  finite quark mass effects are
large, and enhance the  $D_{s0} \to D^*_s \gamma$ amplitude by at least a factor of two.

From the second sum rule eq.(\ref{eq:d1-hqet-1}),  taking
into account the  dependence on the Borel parameter $E$ for the continuum
subtraction, we obtain a smaller result, see fig. \ref{fig:d1-hqet} (right). This is due to a nearly complete cancellation between two different terms, the perturbative and the leading twist term, and therefore
it critically depends on the input parameters of the QCD side of the sum rule,
 making the numerical result less reliable.

\section{$D_{sJ}(2460) \to D_s \gamma$} \label{sec:Ds1a}
Coming back to the case of finite charm quark mass,  let us
consider   three radiative decay modes of  $D_{s1}^\prime$,  the transitions into
a pseudoscalar $D_s$, a vector $D_s^*$ and a scalar $D_{s0}$ meson with the emission of a photon.  
The calculation of the decay amplitudes  is   analogous to the one carried out in Section \ref{sec:Ds0},  therefore
we present  only the relevant formulae.

The  decay amplitude of $D_{s1}^\prime \to D_s \gamma$:
\be
\langle \gamma(q,\lambda) D_s(p)| D_{s1}^\prime(p+q,\lambda^{\prime\prime})\rangle =  e g_1  \left[ (\varepsilon^* \cdot  \eta)(p \cdot q)-(\varepsilon^* \cdot p)(\eta \cdot q) \right]   \label{eq:ampDs1a}
\ee
with  $\eta(\lambda^{\prime\prime} )$ the  $D_{s1}^\prime$ polarization vector,
involves the hadronic parameter $g_1$ that can be computed  considering the correlation function
\be
T_\mu(p,q)=i \int d^4x \; e^{i p \cdot x} \langle \gamma(q,\lambda) | T[J^\dagger_5(x) J^A_\mu(0)]|0\rangle \,\,\, .
\label{eq:corr-Ds1Dsgamma}
\ee
The quark currents are $J_5=\bar c i \gamma_5 s$ and $J^A_\mu=\bar c \gamma_\mu \gamma_5 s$;  $T_\mu$  can be expanded  in Lorentz-invariant structures:
\be
T_\mu(p,q)= T \;   \left[  (\varepsilon^* \cdot p)  q_\mu - (p\cdot q) \varepsilon^* _\mu  \right]   + ... \,\,\, .
\ee
The sum rule for  $g_1$, obtained from the function $T$, reads:
\bea
g_1&=& \frac{e^{m^2_{D_{s1}^\prime}+m^2_{D_s}\over 2 M^2 } (m_c+m_s)}{m_{D_{s1}^\prime} f_{D_{s1}^\prime}  m^2_{D_s} f_{D_s}} \Bigg\{ \int_{(m_c+m_s)^2}^{s_0} d s  \; e^{- {s \over M^2}} \rho^P(s) \nn\\
&+& e_c \; e^{-{m_c^2 \over M^2}} \ss \left[1-{m_c m_s \over M^2} +{m_s^2 \over 2 M^2}\left(1+{m_c^2 \over M^2}\right) \right]   \nn \\
&-& e_s \ss (e^{-{m_c^2 \over M^2}}- e^{-{s_0 \over M^2}}) M^2 \chi \phi_\gamma(u_0) \nn\\
&-& e_s \ss e^{-{m_c^2 \over M^2}} \Bigg[ -\frac{1}{4} \left(\mathbb{A}(u_0)-8 \bar H_\gamma(u_0)\right)\left(1+{m_c^2\over M^2}\right)\nn\\
&-&\int_0^{1-u_0} dv \int_0^{u_0 \over 1-v} d \alpha_g  {\cal F}_1(u_0-(1-v)\alpha_g, 1-u_0-v \alpha_g,\alpha_g)\nn\\
&-&\int^1_{1-u_0} d v \int_0^{1-u_0 \over v} d \alpha_g  {\cal F}_1(u_0-(1-v)\alpha_g, 1-u_0-v \alpha_g,\alpha_g)\Bigg]\nn\\
&+& 2 e_s f_{3 \gamma} m_c e^{-{m_c^2 \over M^2}}  \Psi^v(u_0) \Bigg\}  \,\,\, , \label{eq:g1}
\eea
where
${\cal F}_1= {\cal S} + \tilde{\cal S}-T_1-T_2+T_3+T_4+2 v (- {\cal S}-T_3+T_2)$
and the  spectral function $\rho^P$ is:
\bea
\rho^P(s)&=&- {3 e_s \over 8 \pi^2} \Bigg\{ 2 m_s \ln \left({s-m_c^2+m_s^2-\lambda^{1\over 2}(s, m_c^2,m_s^2) \over s-m_c^2+m_s^2+\lambda^{1\over 2}(s, m_c^2,m_s^2) } \right) \nn \\
&+&(m_c-m_s){(m_c^2-m_s^2-s) \over s^2}  \lambda^{1\over 2}(s, m_c^2,m_s^2) \Bigg\} \,\, - \,\,(s \leftrightarrow c) \,\,\, .
\eea
Eq.(\ref{eq:g1}) involves  parameters already used in previous Sections and the photon DA
collected in the Appendix; it also involves  the leptonic constant  $f_{D_{s1}^\prime}$ defined by
the matrix element
\be
\langle 0 | J^A_\mu | D_{s1}^\prime \rangle = f_{D_{s1}^\prime}  m_{D_{s1}^\prime} \eta_\mu \,\,\, ,
\ee
which can be obtained, starting from the two-point correlation function
\be
\Pi_{\mu \nu}(p^2)=i \int d^4x \; e^{i p \cdot x} \langle 0 | T[J^A_\mu(0) J^{A \dagger}_\nu(x)] |0\rangle \,\,\, ,
\label{eq:corr-fDs1}
\ee  
from the sum rule:
\bea
 f_{D_{s1}^\prime}^2&=&\frac{e^{m^2_{D_{s1}^\prime}\over  M^2 }}{m_{D_{s1}^\prime}^2} \Bigg\{\frac{1}{8 \pi^2} \int_{(m_c+m_s)^2}^{s_0} ds \; \lambda^{1\over2}(s,m_c^2,m_s^2)\left[2-{m_c^2+m_s^2+6 m_c m_s\over s} -{(m_c^2-m_s^2)^2\over s^2}      \right] e^{-{s\over M^2}} \nn\\
 &+&e^{-{m^2_c \over  M^2}}\left[ \ss \left(m_c -{m_c^2 m_s \over 2 M^2}+{m_c^3 m_s^2 \over 2 M^4}\right)-{\mixedss\over 4}{m_c^3\over M^4}\right]\Bigg\} \,\,\, .
\eea
We get  $f_{D_{s1}^\prime} \simeq f_{D_{s0}}$. 

\begin{figure}[h]
 \begin{center}
  \includegraphics[width=0.5\textwidth] {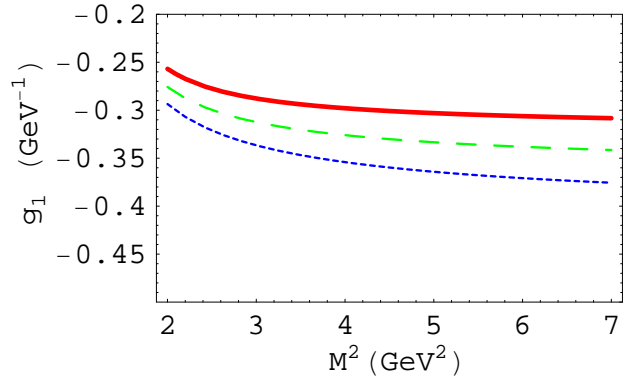}
\vspace*{0mm}
 \caption{\baselineskip=15pt 
 The parameter $g_1$ in the $D_{s1}^\prime \to D_s^* \gamma$ decay amplitude, 
 eq.(\ref{eq:ampDs1a}), as a function of  the Borel parameter $M^2$. The  curves refer to 
 the threshold $s_0=2.5^2$ ${\rm GeV}^2$ 
(continuous),  $s_0=2.55^2$ $\rm{GeV}^2$ (long-dashed)  and $s_0=2.6^2$ $\rm{GeV}^2$ 
(dashed line).} 
  \label{fig:g1}
 \end{center}
\end{figure}

The calculation of  $g_1$  produces  the curves depicted in fig.\ref{fig:g1}. 
%
%
Considering the range $3 \,\, {\rm GeV}^2 \le M^2 \le 6 \,\,{\rm GeV}^2$, together with the variation of the threshold
$s_0$, we obtain: $ -0.37 \,\, {\rm GeV}^{-1} \le g_1 \le -0.29 \,\, {\rm GeV}^{-1}$, and therefore
\be
\Gamma(D_{s1}^\prime\to D_s \gamma)= \, (19\, - \, 29) \, \, {\rm keV}.
\ee
\noindent
As for $D_{s0}\to D_s^* \gamma$,  the result  of light-cone sum rules for the  width of
$D_{s1}^\prime\to D_s \gamma$
is larger than  previous estimates. We shall discuss this point later on.

\section{$D_{sJ}(2460) \to D_s^* \gamma$} \label{sec:Ds1b}
\begin{figure}[h]
 \begin{center}
  \includegraphics[width=0.5\textwidth] {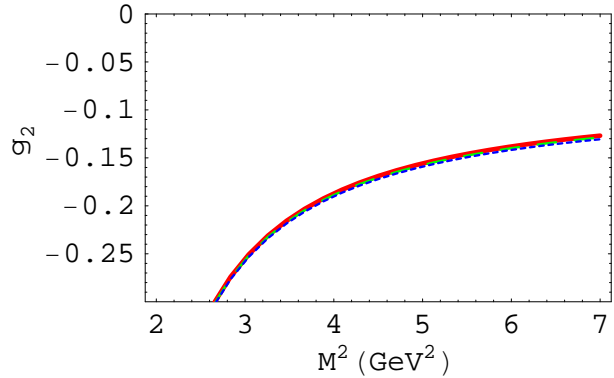}
\vspace*{0mm}
 \caption{\baselineskip=15pt 
The parameter $g_2$ in the  $D_{s1}^\prime \to D_s^* \gamma$  amplitude eq.(\ref{eq:ampDs1b}) versus the Borel parameter $M^2$.  The continuous, long-dashed and  dashed lines refer to $s_0=2.5^2$ ${\rm GeV}^2$,  $s_0=2.55^2$ ${\rm GeV}^2$   and $s_0=2.6^2$ ${\rm GeV}^2$,  respectively.} 
  \label{fig:g2}
 \end{center}
\end{figure}

\begin{figure}[h]
 \begin{center}
  \includegraphics[width=0.5\textwidth] {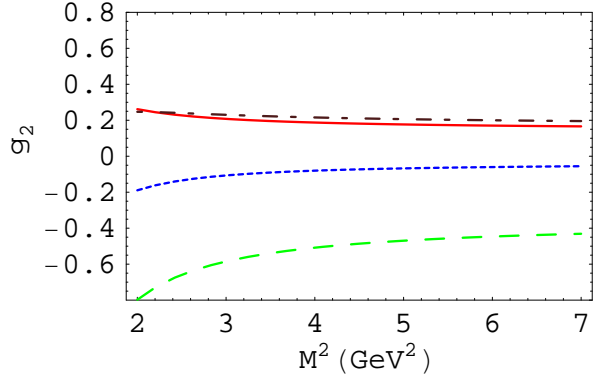}
\vspace*{0mm}
 \caption{\baselineskip=15pt 
 Contributions to the sum rule (\ref{eq:g2}) for  $g_2$.  The  continuous
line corresponds to the perturbative contribution in fig.2 (a,b), the  long-dashed line to the term
proportional to the magnetic  susceptibility of the quark condensate $\chi$,
the  long-short dashed line to the  contribution proportional to $f_{3\gamma}$ and 
the   dashed line to the contribution corresponding to fig.2 (c).
The  threshold is fixed to  $s_0=2.55^2$ ${\rm GeV}^2$ .} 
  \label{fig:g2twist}
 \end{center}
\end{figure}

The calculation of  the dimensionless hadronic parameter $g_2$ appearing in the 
$D^\prime_{s1} \to D_s^* \gamma$ transition amplitude:
\be
\langle \gamma(q,\lambda) D^*_s(p,\lambda^\prime)| D_{s1}^\prime(p+q,\lambda^{\prime \prime})\rangle = i \, e \, g_2 \,  \varepsilon_{\alpha \beta \sigma \tau}  \eta^\alpha \tilde \eta^{*\beta} \varepsilon^{*\sigma}   q^\tau  \,\,\, , \label{eq:ampDs1b}
\ee
with  $\tilde \eta(\lambda^\prime)$  and $\eta(\lambda^{\prime \prime})$ the polarization vectors of  $D^*_s$ and $D_{s1}^\prime$,  is based on the analysis of the correlation function
\be
T_{\mu \nu}(p,q)=i \int d^4x \; e^{i p \cdot x} \langle \gamma(q,\lambda) | T[J^\dagger_\mu(x) J^A_\nu(0)] |0\rangle
\label{eq:corr-Ds1Dstarsgamma}
\ee
expanded in Lorentz invariant structures
\bea
T_{\mu \nu}(p,q)&=& T_A \;   \varepsilon_{\mu \nu \sigma \tau} \varepsilon^{* \sigma} q^\tau +
T_B \;  p_\mu  \varepsilon_{\nu \beta \sigma \tau} p^\beta \varepsilon^{* \sigma} q^\tau \nn\\
&+&T_C \;   (p+q)_\nu \varepsilon_{\alpha \mu  \sigma \tau} p^\alpha \varepsilon^{* \sigma} q^\tau + \dots \,\,\, .
\eea
A sum rule for  $g_2$ is obtained from  $T_A$:
\bea
g_2&=& \frac{e^{m^2_{D_{s1}^\prime}+m^2_{D^*_s}\over 2 M^2 }}{m_{D_{s1}^\prime} f_{D_{s1}^\prime}  m_{D^*_s} f_{D^*_s}} \Bigg\{ \int_{(m_c+m_s)^2}^{s_0} d s  \; e^{- {s \over M^2}} \rho^P(s) 
+ e_c  m_c \; e^{-{m_c^2 \over M^2}} \ss \left[1-{m_s^2 \over  M^2}\left(1-{m_c^2 \over M^2}\right) \right]   \nn \\
&+& e_s m_c \ss (e^{-{m_c^2 \over M^2}}- e^{-{s_0 \over M^2}}) M^2 \chi \phi_\gamma(u_0)  \nn\\
&+& e_s m_c \ss e^{-{m_c^2 \over M^2}} \Bigg[ -\frac{1}{4} {m_c^2 \over M^2}  \mathbb{A}(u_0) -H_\gamma(u_0)(1-u_0)- \bar H_\gamma(u_0) \left(1- \frac{2 m_c^2}{M^2}\right)\Bigg]\nn\\
&+& e_s f_{3 \gamma}  M^2 (e^{-{m_c^2 \over M^2}} - e^{-{s_0 \over M^2}}) \Bigg[ \frac{1}{4} (1-u_0) \psi^{\prime a} (u_0) -
 \frac{1}{4}  \psi^a (u_0)-   \Psi^v (u_0) \left(1+\frac{2 m_c^2}{M^2}\right)\nn\\ 
 &+&(1-u_0)   \psi^v (u_0) \Bigg] \nn\\
 &+& m_c e_s \ss e^{-{m_c^2 \over M^2}} \Bigg[ 
 \int_0^{1-u_0} dv \int_0^{u_0 \over 1-v} d \alpha_g  {\cal F}_2 (u_0-(1-v)\alpha_g, 1-u_0-v \alpha_g,\alpha_g)\nn\\
&+&\int^1_{1-u_0} d v \int_0^{1-u_0 \over v} d \alpha_g  {\cal F}_2 (u_0-(1-v)\alpha_g, 1-u_0-v \alpha_g,\alpha_g)\Bigg]\nn\\
&-&  e_s f_{3 \gamma} M^2 (e^{-{m_c^2 \over M^2}} - e^{-{s_0 \over M^2}}) \Bigg[ 
 \int_0^{u_0} d \alpha_{\bar q}  \,\,\,  \int_{u_0-\alpha_{\bar q}}^{1-  \alpha_{\bar q}}  \frac{d \alpha_g}{\alpha_g^2}\, {\cal F}_3 (1- \alpha_{\bar q} -\alpha_g, \alpha_{\bar q} ,\alpha_g)\nn\\
&-&\int_0^{u_0} d  \alpha_{\bar q} \frac{1}{u_0-\alpha_{\bar q}} \,\,\, {\cal F}_3 (1-u_0,  \alpha_{\bar q},u_0-\alpha_{\bar q})  \Bigg]\Bigg\} \,\,\, , \label{eq:g2}
\eea
with
${\cal F}_2= {\cal S} + \tilde{\cal S}+T_1-T_2-T_3+T_4$ and
$\displaystyle {\cal F}_3= {\cal A} + {\cal V}$.
 The perturbative spectral function $\rho^P$ reads:
\be
\rho^P(s)= {3 e_s \over 4 \pi^2}  m_s m_c \ln \left({s-m_c^2+m_s^2-\lambda^{1\over 2}(s, m_c^2,m_s^2) \over s-m_c^2+m_s^2+\lambda^{1\over 2}(s, m_c^2,m_s^2) } \right) 
 \,\, + \,\,(s \leftrightarrow c) \,\,\, .
\ee
The result  is reported in fig.\ref{fig:g2}.
%
%
%
Considering the range $4 \,\, {\rm GeV}^2 \le M^2 \le 6 \,\,{\rm GeV}^2$ and the variation of the threshold
$s_0$, we get $ -0.18\,\,  \le g_2 \le -0.13 $, i.e.
\be
\Gamma(D_{s1}^\prime\to D_s^* \gamma)= \, (0.6\, - \, 1.1) \, \, {\rm keV}.
\ee
The small value  of $g_2$ is due to  large  cancellations between the
various contributions to the sum rule (\ref{eq:g2}):   perturbative, twist two and higher  twist contributions,
as shown in fig.\ref{fig:g2twist}. In particular, the   contribution proportional to $f_{3\gamma}$
 turns out to be $50\%$  of the  contribution proportional to the magnetic susceptibility of the quark
 condensate.
  In  the cancellation  the detailed shapes
 of the distribution amplitudes and the numerical values of the parameters  are of critical importance;   
 this sensitivity  induces us to consider the  result for $g_2$ as less accurate than the results for  the other channels.    

%

\section{$D_{sJ}(2460) \to D_{sJ}^*(2317) \gamma$} \label{sec:Ds1c}

The last radiative decay mode we consider for $D_{sJ}(2460)$ is the M1 transition  $D^\prime_{s1} \to D_{s0} \gamma$ which is governed by the  amplitude
\be
\langle \gamma(q,\lambda) D_{s0} (p)| D^\prime_{s1}(p+q,\lambda^{\prime \prime})\rangle = i \, e \,  g_3  \, \varepsilon_{\alpha \beta \sigma \tau}  \varepsilon^{*\alpha} \eta^\beta    p^\sigma q^\tau \,\,\,.
\label{eq:ampDs1Ds0}
\ee
The parameter $g_3$ 
can be evaluated starting from   the correlation function
\be
W_\mu(p,q)=i \int d^4x \; e^{i p \cdot x} \langle \gamma(q,\lambda) | T[J^\dagger_0(x) J^A_\mu(0)] |0\rangle
\label{eq:corr-Ds1Ds0gamma}
\ee
written as 
\be
W_\mu=  i \; \varepsilon_{\mu \alpha  \sigma \tau} \varepsilon^{* \alpha} p^\sigma q^\tau\,\,\, W_0 \,\,\, .
\ee
We work out  the sum rule for  $g_3$:
\bea
g_3&=& \frac{e^{m^2_{D^\prime_{s1}}+m^2_{D_{s0}}\over 2 M^2 }}{m_{D_{s0}} f_{D_{s0}}  m_{D^\prime_{s1}} f_{D^\prime_{s1}}} \Bigg\{ \int_{(m_c+m_s)^2}^{s_0} d s  \; e^{- {s \over M^2}} \rho^P(s)  
+ e_c \; e^{-{m_c^2 \over M^2}}  \ss \left(1+{m_s m_c \over 2 M^2}+{m_s^2 m_c^2 \over 8 M^4}\right)  \nn \\
&+& e_s \ss (e^{-{m_c^2 \over M^2}} - e^{-{s_0 \over M^2}} )
M^2 \chi \phi_\gamma(u_0) \nn\\
&+& e^{-{m_c^2 \over M^2}}  e_s \ss [- \frac{1}{4} \mathbb{A}(u_0) (1+\frac{m_c^2}{M^2})]-\frac{m_c}{2} e_s f_{3 \gamma} \psi^a(u_0)  e^{-{m_c^2 \over M^2}}  \nn\\
&+&e^{-{m_c^2 \over M^2}} e_s \ss
\Big[\int^{1-u_0}_0 d v \int_0^{u_0 \over 1- v} d \alpha_g  {\cal F}_4(u_0-(1-v)\alpha_g, 1-u_0-v \alpha_g,\alpha_g) \nn \\
&+& \int^1_{1-u_0} d v \int_0^{1-u_0 \over v} d \alpha_g  {\cal F}_4(u_0-(1-v)\alpha_g, 1-u_0-v \alpha_g,\alpha_g) \Big]
 \Bigg\} \nn\\
\eea
with
${\cal F}_4= {\cal S} + \tilde{\cal S}+T_1+T_4-T_2-T_3+2 v (- \tilde{\cal S}+T_3-T_4)$ and 
\bea
\rho^P(s)&=& {3 e_s \over 4 \pi^2} \left\{ \frac{(m_c+m_s)}{s} \lambda^{1\over 2}(s, m_c^2,m_s^2)  + m_s \ln \left({s-m_c^2+m_s^2-\lambda^{1\over 2}(s, m_c^2,m_s^2) \over s-m_c^2+m_s^2+\lambda^{1\over 2}(s, m_c^2,m_s^2) } \right) \right\} \nn \\
&-& (s \leftrightarrow c) \,\, .
\eea
\begin{figure}[h]
 \begin{center}
  \includegraphics[width=0.5\textwidth] {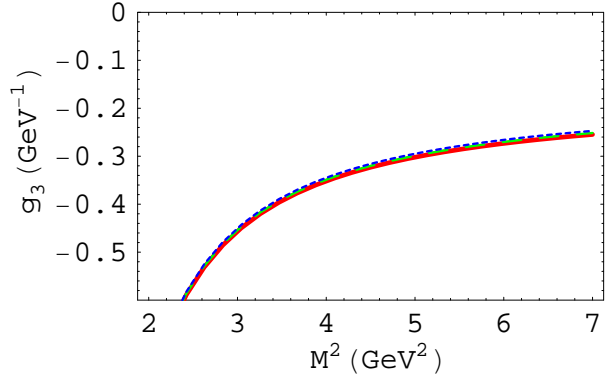}
\vspace*{0mm}
 \caption{\baselineskip=15pt 
The parameter  $g_3$  in the $D_{s1}^\prime \to D_{s0} \gamma$  amplitude, eq.(\ref{eq:ampDs1Ds0}),
 versus the Borel parameter $M^2$.
 The curves correspond to the same thresholds as in fig.  \ref{fig:g1}  and \ref{fig:g2}.} 
  \label{fig:g3}
 \end{center}
\end{figure}
%
Considering the range $4 \,\, {\rm GeV}^2 \le M^2 \le 6 \,\,{\rm GeV}^2$  and  varying   the threshold
$s_0$ we get:  $ -0.35 \,\, {\rm GeV}^{-1} \le g_3 \le -0.27 \,\, {\rm GeV}^{-1}$, corresponding to
\be
\Gamma(D_{s1}^\prime\to D_{s0} \gamma)= \, (0.5\, - \, 0.8) \, \, {\rm keV}.
\ee

\section{Discussion and Conclusions}\label{sec:conc}

As seen in the  previous Sections, the radiative decay amplitudes of the charmed mesons 
considered here, when evaluated by
light-cone QCD sum rules,  are determined by   two main contributions,  the perturbative photon emission from the heavy and light quarks, and the contribution of the photon
emission from the soft light quark.  Other terms represent small corrections. In general, these two
terms have different signs, and produce large cancellations; this allows to understand the
role  of  QCD parameters like the  magnetic susceptibility $\chi$. 
The delicate balancing of the two contributions determines the difference between
the radiative widths of charged and neutral mesons. 

In Table \ref{predictions} we  collect  the LCSR  results 
together with  the results  of  other methods   
\cite{Colangelo:2004vu,Colangelo:2003vg,Godfrey:2003kg,Bardeen}.
%
\begin{table}[h]
\caption{\baselineskip 15pt
Radiative decay widths (in keV) of $D^*_{sJ}(2317)$ and $D_{sJ}(2460)$ obtained by light-cone sum rules
(LCSR).   Vector Meson Dominance (VMD) and constituent quark model (QM) results
are also reported.} \label{predictions}
    \begin{center}
    \begin{tabular}{c c c c c c}
\hline Initial state & Final state & LCSR &VMD \cite{Colangelo:2004vu,Colangelo:2003vg}
& QM \cite{Godfrey:2003kg}  & QM \cite{Bardeen} \\ \hline 
$D^*_{sJ}(2317)$&$D_{s}^{\ast }\gamma$&  4-6     & 0.85& 1.9 &  1.74\\ 
$D_{sJ}(2460)$   &$D_{s}\gamma$             & 19-29 & 3.3&  6.2 & 5.08 \\
                              &$D_{s}^*\gamma$          & 0.6-1.1 & 1.5&  5.5 & 4.66 \\
                               &\,\,\,\,\,$D^*_{sJ}(2317)\gamma$\,\,\,\,\,& 0.5-0.8  &  \ \hfill --- \hfill\ &0.012 & 2.74\\ \hline
\end{tabular}
\end{center}
\end{table}
%
With the  exception of $D_{sJ}(2460)\to D^*_s \gamma$, the rates of all the 
modes  are  larger than  obtained by other approaches. In particular,  $\Gamma(D_{sJ}(2460)\to D_s \gamma)$ turns out to be considerably
wider;  notice that  $D_{sJ}(2460)\to D_s \gamma$
is the only radiative mode observed   so far,  as shown  in Table  \ref{br}.  
The peculiarity 
in  $D_{sJ}(2460)\to D_s \gamma$ is that the perturbative contribution
to the sum rule is  the largest term,  while  in the other cases the leading twist term is the largest one in 
the theoretical side of the sum rules. 
%
\begin{table}[h]
\caption{\baselineskip 15pt
Measurements and 90\% CL upper limits of  ratios of  $D^*_{sJ}(2317)$  and  $D_{sJ}(2460)$  decay widths.} \label{br}
    \begin{center}
    \begin{tabular}{cccc}
\hline & Belle & BaBar   & CLEO \cite{Besson:2003cp} \\ \hline $\frac{\Gamma \left(
D^*_{sJ}(2317) \rightarrow D_{s}^{\ast }\gamma \right) }{ \Gamma \left(
D^*_{sJ}(2317)\rightarrow D_{s}\pi ^{0}\right) }$ & $<0.18$ \cite{Bellecontinuo}&  \ \hfill --- \hfill\   & $<0.059$ \\
\hline $\frac{\Gamma \left(D_{sJ}(2460) \rightarrow
D_{s}\gamma \right) }{ \Gamma \left( D_{sJ}(2460)\rightarrow
D_{s}^{\ast }\pi ^{0}\right) }$ &
\begin{tabular}{c}
$0.55\pm0.13\pm0.08$ \cite{Bellecontinuo}\\
$0.38\pm0.11\pm0.04$ \cite{BelledalB}\\
\end{tabular}&
\begin{tabular}{c}
$0.375\pm0.054\pm0.057$ \cite{BaBardalB} \\
$0.274\pm0.045\pm0.020$  \cite{Aubert:2004pw}
\end{tabular}
&$<0.49$ \\
\hline $\frac{\Gamma \left( D_{sJ}(2460)\rightarrow
D_{s}^{\ast }\gamma \right) }{\Gamma \left( D_{sJ}(2460)\rightarrow D_{s}^{\ast }\pi ^{0}\right) }$ & $<0.31$ \cite{Bellecontinuo}&\ \hfill --- \hfill\ &$<0.16$ \\
\hline  $\frac{\Gamma \left( D_{sJ}(2460)\rightarrow
D^*_{sJ}(2317)\gamma \right) }{\Gamma \left( D_{sJ}(2460)\rightarrow
D_{s}^*\pi^0 \right) }$  & \ \hfill --- \hfill\ & $ < 0.23$   \cite{BaBardalB}& $ < 0.58$\\
\hline
\end{tabular}
\end{center}
\end{table}
%
%

In order to quantitatively understand the data in Table \ref{br} one should precisely know
the  widths of the isospin violating transitions
$D_{s0}\to D_s \pi^0$ and $D^\prime_{s1}\to D_s^* \pi^0$. In the description of    
these  transitions based on the mechanism of $\eta-\pi^0$ mixing 
\cite{wise,close,Godfrey:2003kg,Colangelo:2003vg}
one should  accurately determine  the
strong  couplings $D_{s0} D_s \eta$ and $D^\prime_{s0} D^*_s \eta$  
for finite heavy quark mass.  Considering  the results  in Tables \ref{predictions},\ref{br},
 these couplings should be  larger 
than  obtained in the heavy quark and  SU(3) limit, an issue which 
 deserves further investigation.
 
The obtained  dominance of $D^\prime_{s1}\to D_{s} \gamma$ with respect to other modes,
in agreement with observation, induces us  to consider  our results as consistent  with 
the  interpretation of  $D^*_{sJ}(2317)$ and  $D_{sJ}(2460)$
as ordinary $\bar c s$ mesons.  The  
$D_{sJ}(2460)\to D^*_{sJ}(2317) \gamma$ decay turns out to be suppressed;  in the molecular
interpretation it would be  somehow enhanced. 
The observation of all
the radiative decay modes with the predicted rates  would of course 
reinforce our statement;  in the meanwhile,  we can reasonably conclude  that invoking non-standard
interpretations is not necessary.

\vspace*{1cm}
\noindent {\bf Acnowledgments.}
We thank T. Aliev  for discussions and for collaboration in the early stage of this work.
We acknowledge partial support from the EC Contract No.
HPRN-CT-2002-00311 (EURIDICE).

\newpage
\appendix
\section{Photon distribution amplitudes}\label{app}

For completeness,   we collect  in this Appendix the light-cone expansions of the  photon matrix
elements relevant for the calculation of the radiative decays of $D_{s0}$ and $D^\prime_{s1}$.
We also collect  the expressions of the photon distribution amplitudes and the numerical values of the related parameters, as reported in  \cite{Ball:2002ps}. In all the expressions
 $\varepsilon(\lambda)$ is the photon polarization vector and 
 $\dd \tilde \varepsilon_{\mu}=\varepsilon_\mu^*-q_\mu {\varepsilon^* \cdot x \over q \cdot x} $, 
 $\dd \tilde g_{\mu \nu}=g_{\mu\nu} -\frac{1}{q \cdot x}( q_\mu x_\nu+q_\nu x_\mu)$; 
 the variable $\bar u$ is defined as $\bar u=1-u$;  $\tilde G_{\mu \nu}$ is the dual field
  $\tilde G_{\mu \nu}= \frac{1}{2} \varepsilon_{\mu \nu \alpha \beta}  G^{\alpha \beta}$.
  We neglect quark mass corrections, that have not been worked out for all matrix elements.
\bea
\langle \gamma(q,\lambda)|\bar q(x) \sigma_{\mu\nu} q(0)| 0\rangle&=& -i e e_q \qq
(\varepsilon_\mu^* q_\nu -\varepsilon_\nu^* q_\mu) \int_0^1 du \; e^{i \bar u q \cdot x}\left(\chi \phi_\gamma(u)+{x^2\over 16} \mathbb{A}(u) \right) \nn\\
&-&i e e_q {\qq \over 2 q  x} (x_\nu \tilde \varepsilon_{\mu} -x_\mu \tilde \varepsilon_{\nu} ) \int_0^1 du \; e^{i \bar u q \cdot x} h_\gamma(u) \nn\\
\langle \gamma(q,\lambda)|\bar q(x) \gamma_\mu q(0)| 0\rangle &=&  e e_q  f_{3 \gamma}  \tilde \varepsilon_{ \mu}  \int_0^1 du \; e^{i \bar u q \cdot x}  \psi^v(u)\nn \\
\langle \gamma(q,\lambda)|\bar q(x) \gamma_\mu \gamma_5 q(0)| 0\rangle &=&  - {1\over 4} e e_q  f_{3 \gamma} \epsilon_{\mu\nu\alpha\beta} \varepsilon^{*\nu} q^\alpha x^\beta  \int_0^1 du \; e^{i \bar u q \cdot x}  \psi^a(u) \nn\\
\langle \gamma(q,\lambda)|\bar q(x) g_s G_{\mu\nu}(vx) q(0)| 0\rangle &=&  -i e e_q \qq (\varepsilon_\mu^* q_\nu -\varepsilon_\nu^* q_\mu)  \int {\cal D}\alpha_i  \; e^{i  (\alpha_{\bar q}+v \alpha_g)  q \cdot x}  {\cal S}(\alpha_i) \nn\\
\langle \gamma(q,\lambda)|\bar q(x) g_s G_{\mu\nu}(vx) i \gamma_5 q(0)| 0\rangle &=&  -i e e_q \qq (\varepsilon_\mu^* q_\nu -\varepsilon_\nu^* q_\mu)  \int {\cal D}\alpha_i  \; e^{i  (\alpha_{\bar q}+v \alpha_g)  q \cdot x}  \tilde{\cal S}(\alpha_i) \nn \\
\langle \gamma(q,\lambda)|\bar q(x) g_s \tilde G_{\mu\nu}(vx) \gamma_\alpha \gamma_5 q(0)| 0\rangle &=&   e e_q  f_{3 \gamma} q_\alpha  (\varepsilon_\mu^* q_\nu -\varepsilon_\nu^* q_\mu)  \int {\cal D}\alpha_i  \; e^{i  (\alpha_{\bar q}+v \alpha_g)  q \cdot x}  {\cal A}(\alpha_i) \nn \\
\langle \gamma(q,\lambda)|\bar q(x) g_s  G_{\mu\nu}(vx) i \gamma_\alpha  q(0)| 0\rangle &=&   e e_q  f_{3 \gamma} q_\alpha  (\varepsilon_\mu^* q_\nu -\varepsilon_\nu^* q_\mu)  \int {\cal D}\alpha_i  \; e^{i  (\alpha_{\bar q}+v \alpha_g)  q \cdot x}  {\cal V}(\alpha_i) \nn 
\eea
\bea
\langle \gamma(q,\lambda) \vert \bar q(x) \sigma_{\alpha \beta} g_s G_{\mu \nu}(v x) q(0) \vert 0 \rangle &=& \nn\\
e e_q \qq \Bigg\{
        \Bigg[ \tilde \varepsilon_{\mu}  \tilde g_{\alpha \nu} q_\beta  -
         \tilde \varepsilon_{\mu} \tilde g_{\beta \nu}  q_\alpha 
		 &-& (\mu \leftrightarrow \nu)  \Bigg]
   \int {\cal D}\alpha_i e^{i (\alpha_{\bar q} + v \alpha_g) q\cdot x} {\cal T}_1(\alpha_i) 
\nonumber \\ 
+  \Bigg[ \tilde \varepsilon_{\alpha} \tilde g_{\mu \beta}   q_\nu 
 -  \tilde \varepsilon_{\alpha}  \tilde g_{\nu \beta}  q_\mu  &-&(\alpha \leftrightarrow \beta)  \Bigg]
     \int {\cal D} \alpha_i e^{i (\alpha_{\bar q} + v \alpha_g) q\cdot x} {\cal T}_2(\alpha_i)
\nonumber \\ &+&
        {(q_\mu x_\nu - q_\nu x_\mu)
		(\varepsilon_\alpha^* q_\beta - \varepsilon_\beta^* q_\alpha)\over q\cdot x}
    \int  {\cal D} \alpha_i e^{i (\alpha_{\bar q} + v \alpha_g) q\cdot x} {\cal T}_3(\alpha_i) 
\nonumber \\ &+&
         {(q_\alpha x_\beta - q_\beta x_\alpha)
		(\varepsilon_\mu^* q_\nu - \varepsilon_\nu^* q_\mu) \over q\cdot x}
     \int  {\cal D} \alpha_i e^{i (\alpha_{\bar q} + v \alpha_g) q\cdot x} {\cal T}_4(\alpha_i)
                         \Bigg\}\nn \\
\eea
\noindent $\alpha_i=\{\alpha_q, \alpha_{\bar q}, \alpha_g\}$ and 
$\dd \int{\cal D}(\alpha_i)  \equiv \int_0^1 d\alpha_q \int_0^1d\alpha_{\bar q} \int_0^1 d\alpha_g \delta (1-\alpha_q-\alpha_{\bar q}- \alpha_g)$. The  photon distribution amplitudes (DA) 
 have the following expressions:
\bea
\phi_\gamma(u)& = & 6 u \bar u \left(1+\varphi_2 C_2^{\frac{3}{2}}(2 u- 1) \right ) \nn \\
\mathbb{A}(u) &=& 40 u^2 \bar u^2 (3k - k^+ + 1) + 8 (\zeta_2^+ - 3\zeta_2)\big[u \bar u (2 + 13u \bar u) \nn\\
&+& 2u^3(10 - 15u + 6u^2) \ln u + 2\bar u^3(10 - 15 \bar u + 6 \bar u^2)\ln \bar u \big]\nn \\
h_\gamma(u)&=&-10 (1 + 2  k^+) C_2^{\frac{1}{2}} (2 u - 1) \nn \\
\psi^v(u)&=& 5\left(3 (2  u - 1)^2 - 1\right) + \frac{3}{64} \left(15  \omega^V_\gamma - 5   \omega^A_\gamma \right) \left(3 - 30 (2  u - 1)^2 + 35 (2  u - 1)^4\right) \nn\\
\psi^a(u)&=&  \left(1-(2  u - 1)^2\right) \left(5(2 u -1)^2-1\right)  \frac{5}{2} \left(1+ \frac{9}{16} \omega^V_\gamma - \frac{3}{16}   \omega^A_\gamma \right) \nn \\
{\cal V}(\alpha_q, \alpha_{\bar q}, \alpha_g) &=& 540 \omega_\gamma^V ( \alpha_q-\alpha_{\bar q}) \alpha_q \alpha_{\bar q} \alpha_g^2 \nn\\
{\cal A}(\alpha_q, \alpha_{\bar q}, \alpha_g) &=& 360 \alpha_q \alpha_{\bar q} \alpha_g^2  \left[ 1+ \omega_\gamma^A \frac{1}{2}(7  \alpha_g-3) \right] \nn 
\eea
\bea
{\cal S}(\alpha_q, \alpha_{\bar q}, \alpha_g) &=&  30 \alpha_g^2 \Big[(k + k^+)(1 - \alpha_g) + (\zeta_1 + \zeta_1^+)(1 - \alpha_g)(1 - 2 \alpha_g) \nn \\
&+&  \zeta_2 [3 (\alpha_{\bar q} - \alpha_q)^2 - \alpha_g(1 - \alpha_g)]\Big] \nn \\
\tilde {\cal S}(\alpha_q, \alpha_{\bar q}, \alpha_g) &=&  - 30 \alpha_g^2 \Big[(k - k^+)(1 - \alpha_g) + (\zeta_1 - \zeta_1^+)(1 - \alpha_g)(1 - 2 \alpha_g) \nn\\
&+& \zeta_2[3(\alpha_{\bar q} - \alpha_q)^2 - \alpha_g(1 - \alpha_g)]\Big] \nn \\
{\cal T}_1 (\alpha_q, \alpha_{\bar q}, \alpha_g) &=&  -120 (3 \zeta_2 + \zeta_2^+)(\alpha_{\bar q} - \alpha_q) \alpha_q \alpha_{\bar q} \alpha_g \nn \\
{\cal T}_2 (\alpha_q, \alpha_{\bar q}, \alpha_g) &=& 30 \alpha_g^2 (\alpha_{\bar q} - \alpha_q)\left[(k - k^+) + (\zeta_1 - \zeta_1^+)(1 - 2 \alpha_g) + \zeta_2(3 - 4 \alpha_g)\right] \nn \\
{\cal T}_3 (\alpha_q, \alpha_{\bar q}, \alpha_g) &=&-120 (3 \zeta_2 - \zeta_2^+)(\alpha_{\bar q} - \alpha_q)\alpha_q  \alpha_{\bar q} \alpha_g \nn \\
{\cal T}_4 (\alpha_q, \alpha_{\bar q}, \alpha_g) &=&  30 \alpha_g^2 (\alpha_{\bar q} - \alpha_q)\left[(k + k^+) + (\zeta_1 + \zeta_1^+)(1 - 2 \alpha_g) + \zeta_2(3 - 4 \alpha_g)\right]   .  \nn \\
\eea
The parameters  in the  distribution amplitudes  are:
$f_{3 \gamma}=-(0.0039\pm0.0020) \, {\rm GeV}^2$, 
$\omega^V_\gamma = 3.8\pm 1.8$, $\omega^A_\gamma = -2.1\pm  1.0$ \cite{Ball:2002ps};
$k=0.2$,  $\zeta_1=0.4$,   $\zeta_2=0.3$,   $\varphi_2=k^+=\zeta_1^+=\zeta_2^+=0$ (at the renormalization scale $\mu=1 $ GeV) \cite{Balitsky:1989ry}. The other parameters in the QCD sides
of the  sum rules, at the same renormalization scale, are:
$m_c=1.35$ GeV,
$m_s=0.125$ GeV \cite{Colangelo:1997uy},
$\langle \bar s s \rangle=0.8 \langle \bar q q \rangle$ ($q=u,d$),
$\langle \bar q q \rangle=(-0.245 \,\, {\rm GeV})^3 $ and $\langle \bar q g \sigma G q \rangle=m_0^2
\langle \bar q q \rangle$ with $m_0^2=0.8$ GeV$^2$.
Finally, for the magnetic susceptibility  of the quark condensate $\chi$ we use the value $\chi=-(3.15 \pm0.3) \,\, {\rm GeV}^{-2} $  obtained in \cite{Ball:2002ps}.

\end{document}